\documentclass[aps,prl,a4paper,10pt,twocolumn,superscriptaddress]{revtex4-1}

\usepackage{graphicx}
\usepackage[multi-part-units = single]{siunitx}
\usepackage{amsmath}

\begin{document}

\title{Continous every-single-shot carrier-envelope phase measurement \\and control at 100\,kHz}

\author{Dominik Hoff}
\affiliation{Institut f\"ur Optik und Quantenelektronik, Friedrich-Schiller-Universit\"at Jena 
and Helmholtz-Institut Jena, Max-Wien-Platz~1, D-07743 Jena, Germany}

\author{Federico J. Furch}
\affiliation{Max-Born-Institute, Max-Born-Strasse 2a, 12489 Berlin, Germany}

\author{Tobias Witting}
\affiliation{Max-Born-Institute, Max-Born-Strasse 2a, 12489 Berlin, Germany}

\author{Klaus R\"uhle}
\affiliation{Institut f\"ur Optik und Quantenelektronik, Friedrich-Schiller-Universit\"at Jena 
and Helmholtz-Institut Jena, Max-Wien-Platz~1, D-07743 Jena, Germany}

\author{Daniel Adolph}
\affiliation{Institut f\"ur Optik und Quantenelektronik, Friedrich-Schiller-Universit\"at Jena 
and Helmholtz-Institut Jena, Max-Wien-Platz~1, D-07743 Jena, Germany}

\author{A. Max Sayler}
\affiliation{Institut f\"ur Optik und Quantenelektronik, Friedrich-Schiller-Universit\"at Jena 
and Helmholtz-Institut Jena, Max-Wien-Platz~1, D-07743 Jena, Germany}

\author{Marc J. J. Vrakking}
\affiliation{Max-Born-Institute, Max-Born-Strasse 2a, 12489 Berlin, Germany}

\author{Gerhard G. Paulus}
\affiliation{Institut f\"ur Optik und Quantenelektronik, Friedrich-Schiller-Universit\"at Jena 
and Helmholtz-Institut Jena, Max-Wien-Platz~1, D-07743 Jena, Germany}

\author{Claus Peter Schulz}
\affiliation{Max-Born-Institute, Max-Born-Strasse 2a, 12489 Berlin, Germany}

\begin{abstract}
With the emergence of high-repetition-rate few-cycle laser pulse amplifiers, aimed at investigating 
ultrafast dynamics in atomic, molecular and solid state science, the need for ever faster 
carrier-envelope phase (CEP) detection and control has arisen. 
Here we demonstrate a high speed, continuous, every-single-shot measurement and fast feedback 
scheme based on a stereo above-threshold ionization time-of-flight spectrometer capable of detecting 
the CEP and pulse duration at a repetition rate of up to 400\,kHz.
This scheme is applied to a 100\,kHz optical parametric chirped pulse amplification (OPCPA) 
few-cycle laser system, demonstrating improved CEP stabilization and allowing for CEP tagging.
\end{abstract}

\maketitle

% \section{Introduction}

In recent years, the investigation of ultrafast processes and, in particular,
applications in attosecond science motivated the development of new types of coherent light 
sources, which are capable of delivering few-cycle pulses at high average power (up to 200\,W) and 
high repetition rate ($\approx 100$\,kHz).
These light sources are based either on optical parametric amplification (OPA)
\cite{Rothhardt2012,Prinz2015,Matyschok2013,Furch2017,Elu2017,Thire2017}
or direct post-compression of laser pulses from high repetition rate laser amplifiers 
\cite{Hadrich2016}.
The high repetition rate is crucial in experimental techniques where the number of events per 
laser shot has to be kept small or even below one.
Examples of this are the detection of charged particles in coincidence upon photoionization 
\cite{Ullrich2003, Long2017}, or time resolved photoelectron spectroscopy in solids 
\cite{Stockman2007}.
Furthermore, ultrashort pulse duration, down to a few optical cycles, is highly attractive for the 
study and control of ultrafast processes in light-matter interaction. 
In particular, the few-cycle pulse duration allows for the production of isolated attosecond 
extreme ultraviolet (XUV) light pulses through high-harmonic generation (HHG). 
In order to confine the harmonic emission to a single XUV pulse, it is essential to precisely 
measure and control the carrier-envelope phase of the ultrashort pulses. 
Thus, for the successful implementation of high repetition rate sources in attosecond science
applications, the CEP of the pulses has to be determined and possibly controlled at high repetition 
rates exceeding 100\,kHz.

Above-threshold ionization (ATI) of xenon is known to be an effective way for determining the CEP 
of few-cycle pulses \cite{Wittmann2009, Sayler2011, Rathje2012}.  
Thus, it has also been utilized for CEP tagging, which allows for the investigation of 
CEP-dependencies in ultrafast processes by tagging the events of interest with the CEP of the laser 
pulse inducing them \cite{Johnson2011, Sayler2015, Hoff2017}.  
To this end a stereo-ATI CEP-meter (CEPM) and adjacent electronics has been developed over the last 
years, proving control over the CEP up to 10\,kHz \cite{Carpeggiani2017}. 
In this letter, we develop electronics, which calculates the CEP within $\approx300$\,ns after the 
laser pulse, to measure the CEP of a 100\,kHz OPCPA laser system \cite{Furch2017}. 
This not only enables continuous every-single-shot measurements up to several hundreds of kHz 
repetition rate, but also facilitates improved CEP stability and CEP tagging.

% \section{Method}

\begin{figure}[htb]
\centering
\includegraphics[width=0.85\linewidth]{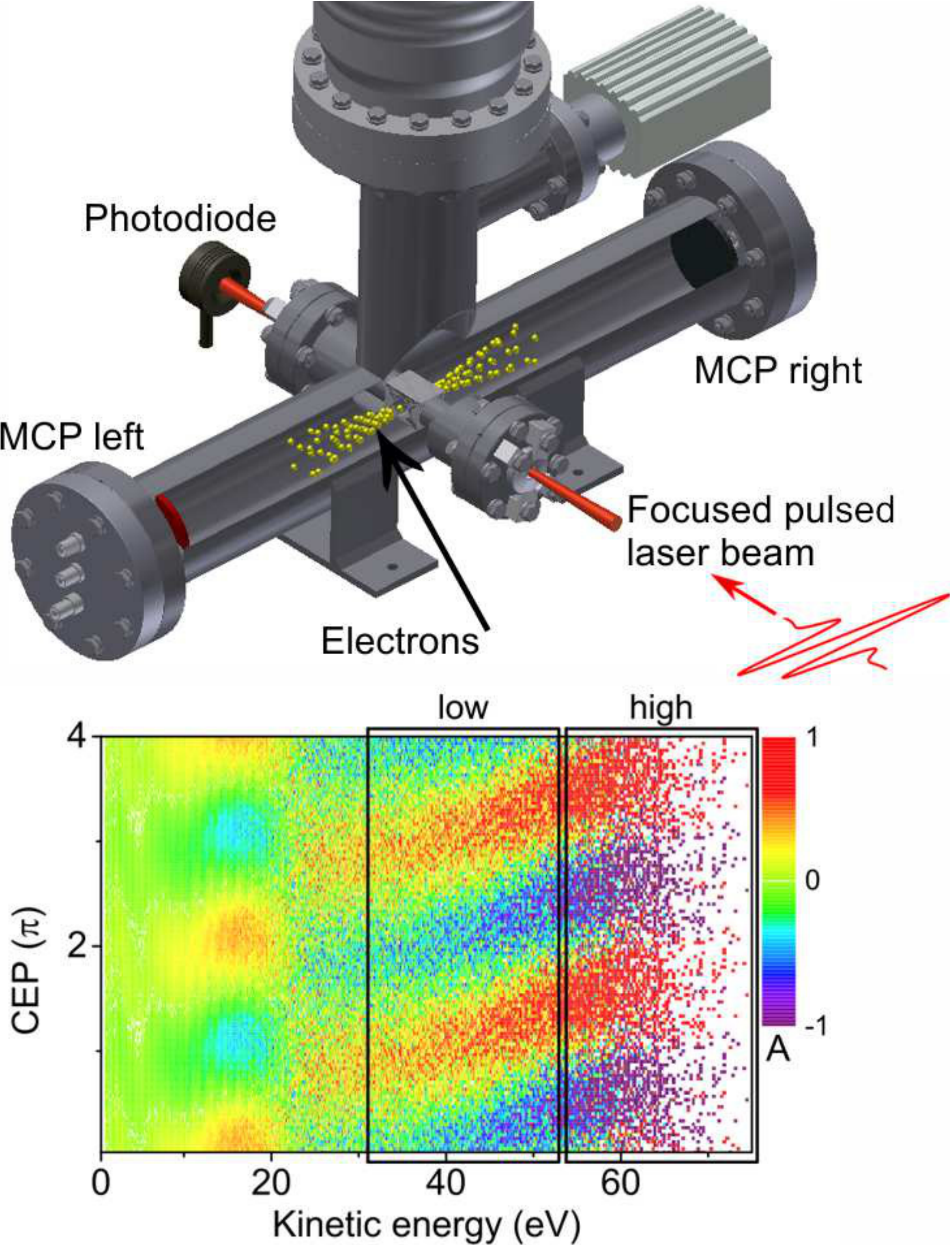}
\caption{Schematic of the stereo ATI apparatus. 
Two TOF spectrometers located on opposite sides along the laser polarization (left/right) measure 
electron spectra of Xe ionized by few-cycle laser pulses. 
The asymmetry (A) in the number of photoelectrons on the left (L) and right (R) detectors, 
$A=(L-R)/(L+R)$, is calculated in two regions in the plateau of the ATI spectra (low/high, bottom 
panel). 
From this, the value of the CEP can be retrieved, see \cite{Wittmann2009, Sayler2011, Sayler2011a, 
Hoff2017}.
}
\label{fig:concept}
\end{figure}

\begin{figure}[ht]
\centering
\includegraphics[width=0.9\linewidth]{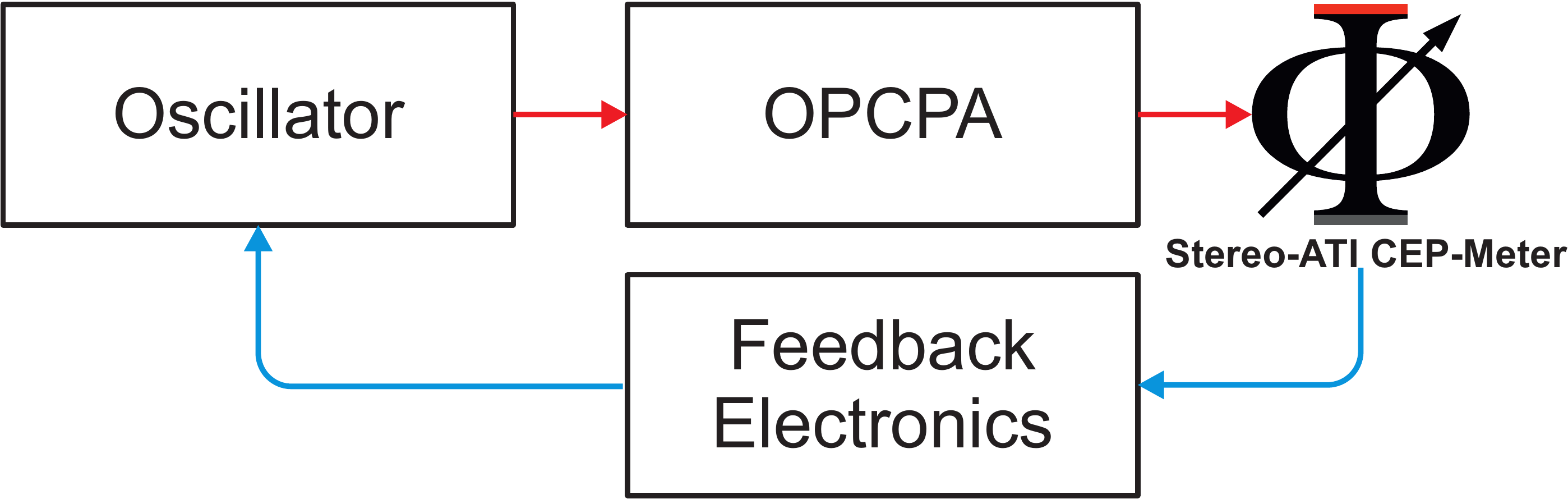}
\caption{
Feedback schematics. 
The oscillator seeds the OPCPA and the amplified pulses are then measured by the CEPM from which a 
feedback signal can be generated and sent to a modulation input in the phase-locking system of the 
oscillator for CEP stabilization up to a bandwidth of 300\,kHz. 
Details about the setup can be found in: oscillator \cite{Rausch2009}, OPCPA \cite{Furch2017} and 
CEPM \cite{Sayler2011}.}
\label{fig:scheme}
\end{figure}

\begin{figure}[htb]
\centering
\includegraphics[width=0.8\linewidth]{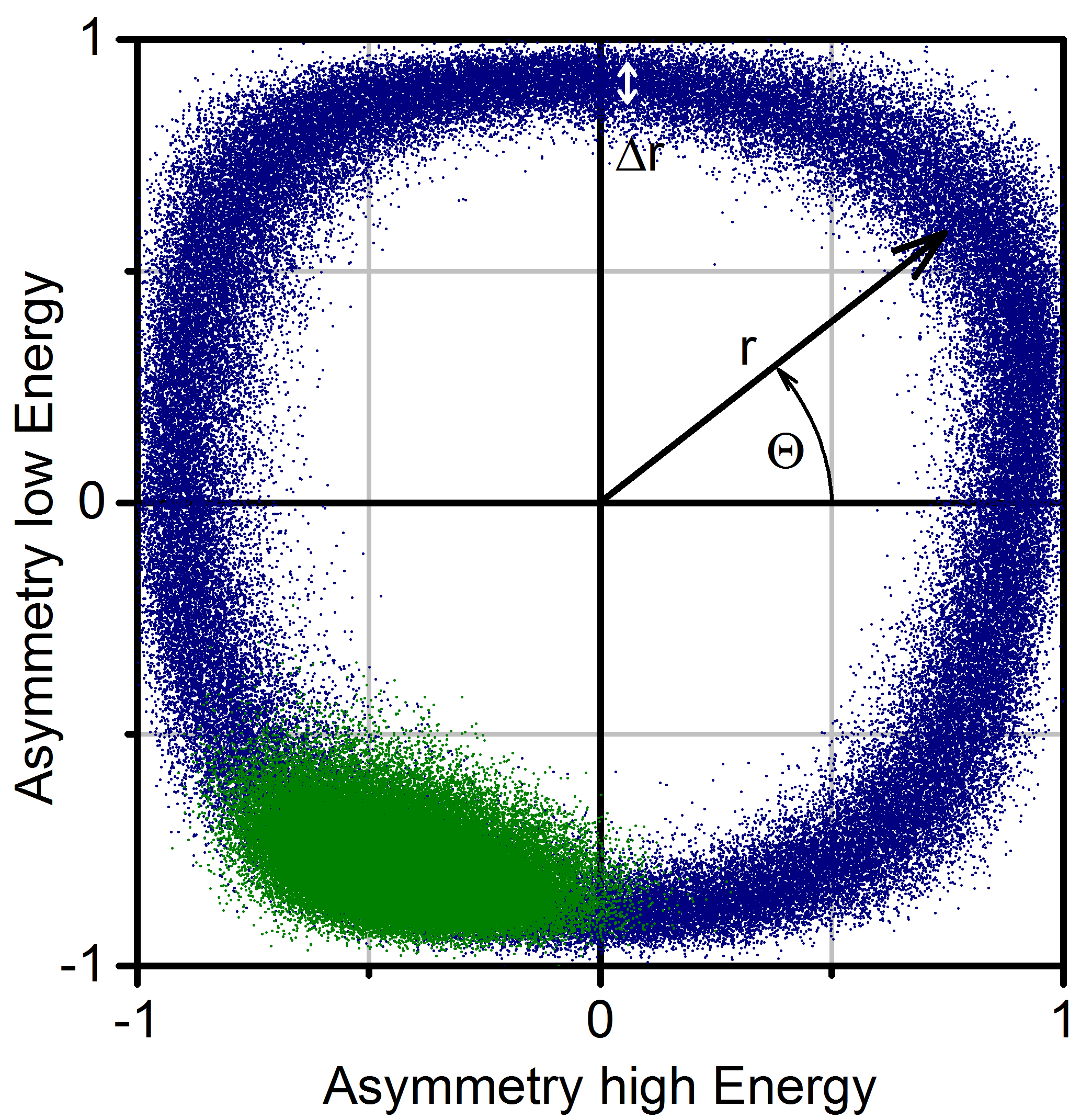}
\caption{Parametric asymmetry plot with $100,000$ individual and consecutive laser shots acquired 
during 1\,s without (blue) and with (green) active CEP-stabilization. 
The CEP is encoded in the polar angle $\Theta$ of the circular shape and the pulse duration in the 
radius $r$, for a detailed description see \cite{Wittmann2009, Sayler2011, Sayler2011a}. 
In this example the average radius corresponds to a pulse duration of 3.5\,fs and the individual 
CEP determinations have an uncertainty of 82\,mrad or 35\,as, estimated from the radius 
distribution $\Delta r$.} 
\label{fig:potato}
\end{figure}

The concept is illustrated in Figs.~\ref{fig:concept} and \ref{fig:scheme}. 
Two time-of-flight (TOF) spectrometers with micro-channel plates (MCP, left and right in 
Fig.~\ref{fig:concept}) face each other along the direction of the laser polarization. 
Given the few-cycle duration of the pulse, there is a difference between the ATI spectrum 
recorded by the two TOF spectrometers (labelled left and right), and this difference is highly 
sensitive to the value of the CEP. 
The left-right asymmetry of the ATI electron spectra is evaluated from two energy regions in the 
plateau of rescattered electrons, as shown in the lower part of Fig. \ref{fig:concept} 
\cite{Wittmann2009}. 
From the asymmetry values of many laser shots, a parametric asymmetry plot can be constructed, Fig. 
\ref{fig:potato}. 
Here the asymmetry of the high-kinetic-energy part of the spectrum of each shot is plotted on the 
x-axis and that of the low kinetic energy part is plotted on the y-axis.
Each and every individual laser shot is measured  and plotted as a single data point, with its CEP 
encoded in the polar angle of the circle. 
Additionally, the pulse duration is encoded in the radius \cite{Sayler2011a}, i.e. the shorter 
the pulse, the larger the asymmetry and the larger the radius.

For the measurement we use the scheme depicted in Fig. \ref{fig:scheme}.
A carrier-envelope offset (CEO) frequency stabilized few-cycle laser oscillator is seeding the 
OPCPA \cite{Rausch2009, Furch2017}. 
The oscillator operates at a repetition rate of 80\,MHz and the CEO frequency is locked to 1/4 of 
the repetition rate. 
As a consequence, the CEP slip from pulse-to-pulse in the pulse train from the oscillator is 
$\pi$/2, and the amplified pulses at 100\,kHz should ideally carry the same value of CEP. 
Thus, for the sake of brevity, we will refer to the oscillator as CEP-stable, when it is operated 
with a locked CEO frequency.
After amplification and compression, 7\,fs, 800\,nm, 190\,$\mu$J pulses are produced at a 
repetition rate of 100\,kHz.
The amplified pulse spectrum was further broadened in a differentially pumped hollow-core fibre 
(HCF) backed with 2.6\,bars of neon gas on the exit side and the pulses were recompressed with 
chirped mirrors and thin fused-silica wedges to an approximate duration of 3.5$\pm$0.5\,fs with a 
pulse energy of $\approx$90\,$\mu$J \cite{Okell2013}.
The pulse duration has been independently confirmed using SEA-F-SPIDER \cite{Witting2011}.

A fraction of 30\,$\mu$J was split off the train and the CEP of the 1.3\,cycle pulses was 
sampled with the stereo-ATI at a rate of 100\,kHz.
The ATI-signal can be evaluated in real time, i.e. within 300\,ns and well before the next pulse 
arrives, and fed back to the oscillator for compensation of ``slow'' CEP drifts in the laser 
system.
In addition, the CEP of the amplified pulses was characterized in parallel utilizing an f-2f 
interferometer \cite{Kakehata2001}. 
In this case each phase value retrieved integrates 100 shots and the detection bandwidth is limited 
to 200\,Hz, limited by the spectrometer utilized. 
These f-2f phase values can alternatively be used to generate a correction signal for feeding back 
to the locking electronics in the oscillator and compensate the slow CEP drifts. 
In this case, the stereo-ATI can be utilized to perform an out-of-loop measurement.

% \section{Results}

Figure \ref{fig:potato} shows a parametric asymmetry plot taken with the 100\,kHz laser system.  
When the oscillator is not CEP-stabilized, i.e. for a randomly varying CEP, the plot approaches a 
circle (blue). 
When the CEP stabilization is switched on, only a section, corresponding to a limited range of CEP 
values, remains (green).

\begin{figure}[htb]
\centering 
\includegraphics[width=\linewidth]{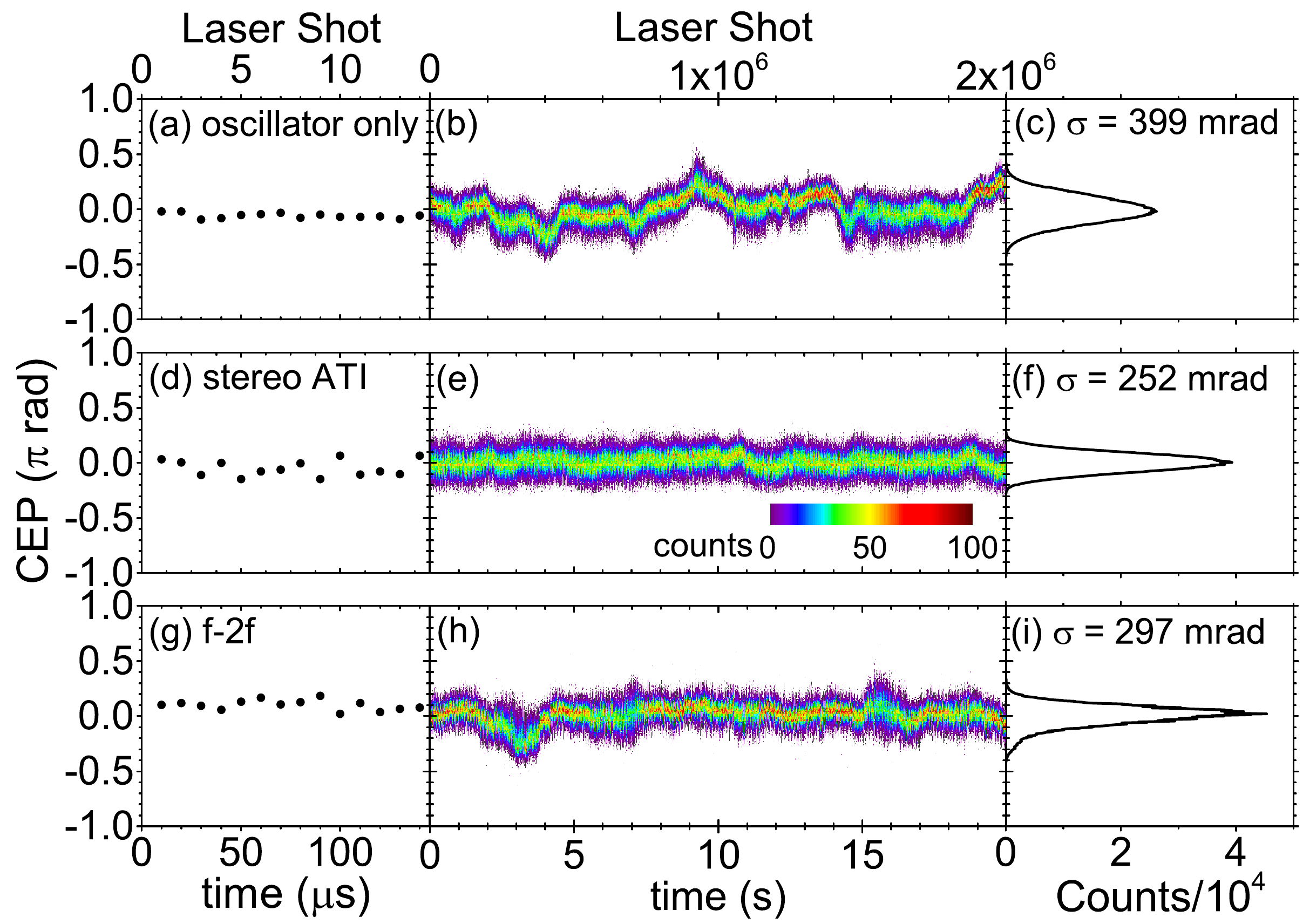}
\caption{Evaluation of three CEP-traces.
Top row (a, b, c) with active oscillator CEP-stabilization only (open loop);
middle row (d, e, f) with Stereo-ATI CEP feedback on the oscillator
and bottom row (g, h, i) f-2f CEP feedback.
The left column shows the short term trace over 200\,$\mu$s and the middle column density plots  
over 20\,s/2M shots with its histogram in the right column. 
For the oscillator-stable only case the SD over the 2M shots amounts to 399\,mrad; while including 
also the Stereo-ATI feedback it is reduced to 252\,mrad, a trace with the f-2f feedback yields 
297\,mrad.}
\label{fig:CEPtrace}
\end{figure}

The performance of the laser system is further characterized in Figure \ref{fig:CEPtrace}, which 
shows the retrieved CEP values as a function of time under different conditions. 
Figure \ref{fig:CEPtrace} (a and b) show the CEP of the amplified pulses when the OPCPA is seeded 
with CEP-stable pulses from the oscillator, but no additional feedback is implemented (open loop).
Given the large number of data points, the CEP values in Fig. \ref{fig:CEPtrace} (b) have been 
arranged in a two-dimensional histogram in the form of a density plot for better visualization.
Figure \ref{fig:CEPtrace} (a) on the other hand, shows the CEP values over the first 200\,$\mu$s, 
in order to highlight the capability of characterizing the CEP of the system on a pulse-to-pulse 
basis. 
This clearly demonstrates that the CEP stability of the oscillator is retained in the OPCPA 
process and mainly low frequency noise remains to be corrected.
Figure \ref{fig:CEPtrace} (c) shows a histogram of the CEP values over the entire series. 
In this open loop case the standard deviation (SD) of the CEP noise over 20\,s, i.e. 
$2\times10^6$\,shots, is 399\,mrad.
Panels (d - f) show similar plots for the case when the stereo-ATI CEP feedback is active (closed 
loop). 
In this case the SD reduces to 252\,mrad and the overall drift is minimized. 
However, from Fig. \ref{fig:CEPtrace} (d), highlighting individual CEP measurements on a short time 
scale, it can be observed that the feedback from the stereo-ATI increases slightly the shot-to-shot 
variations of the CEP. 
This is an indication that there is room for improvement in the feedback loop. 
Finally, Figs. \ref{fig:CEPtrace} (g - h) show similar results for the case when the signal from 
the f-2f interferometer is utilized to implement a closed-loop, and the CEP is characterized 
out-of-loop with the stereo-ATI. 
In this case, the residual CEP noise amounts to 297\,mrad over 20\,s. 
On short time scales the CEP noise does not vary significantly, as in the open-loop case, but the 
variation of the CEP over longer time scales is slightly larger in this case in comparison with Fig. 
\ref{fig:CEPtrace} (e). 

The residual CEP noise values shown in Figs. \ref{fig:CEPtrace} (c, f, i) result from the 
convolution of the residual CEP noise with the response of the CEPM. 
Assuming Gaussian shapes for both, the CEPM response ($\approx$80\,mrad) and the CEP noise, the SD 
excluding CEPM uncertainty for the close-loop measurements are 235\,mrad for the case of the 
stereo-ATI feedback and 284\,mrad for the f-2f case. 
Both results are comparable to values published for typical Ti:Sapphire amplifier systems 
\cite{Koke2008, Canova2009, Adolph2011}, and comply with the demand of attosecond science 
experiments, although not yet reaching the limit imposed by the residual CEP noise in the oscillator 
($\approx$150\,mrad). 
The additional CEP noise in the amplified pulses may have multiple sources, such as pulse-to-pulse 
intensity variations in the pump laser of the OPCPA, residual jitter in the pump-seed delay of the 
OPCPA and beam pointing instabilities affecting the amplification process and the broadening in the 
HCF \cite{Lucking2014, Hadrich2012, Rossi2018}. 

What needs to be emphasized here is that single shot CEP values measured at the full repetition rate
are the basis for the statistics.
So far, the analysis of the CEP residual noise from high repetition rate systems was limited to
measurement of quantities averaged over 10s to 100s of laser shots due to the relatively long 
acquisition times of spectrometers used to analyze interference fringes from f-2f interferometers. 
Thus, the averaged noise values are always significantly smaller than values of single-shot 
based statistics \cite{Rothhardt2012, Matyschok2013, Furch2017}. 
An exception to this averaging are the results published by Thir\'e et al. \cite{Thire2017} 
where a fast detector is utilized to measure single-shot interferograms at 10\,kHz out of a 
100\,kHz pulse train. 
However, undersampling the pulse train can also underestimate the CEP noise, as the 
observed increasing high frequency noise can not be captured.

In contrast, thanks to the fast acquisition rate of the stereo-ATI apparatus, it is possible to 
retrieve the CEP of all the pulses in the 100\,kHz train.
If these single shot CEP values are averaged in groups of 100 shots, as the f-2f measurement does, 
one obtains a SD of just 133\,mrad over 10\,s for the stereo-ATI feedback stabilized case in 
Fig.\ref{fig:CEPtrace} (e). 
This explains and corresponds well to the lower value obtained with an f-2f interferometer 
integrating 100 shots \cite{Furch2017}. 
However, these levels of single shot CEP-noise are well suitable to perform CEP-sensitive 
experiments that rely on individual attosecond pulses \cite{Carpeggiani2017}.

\begin{figure}[htb]
\centering
\includegraphics[width=\linewidth]{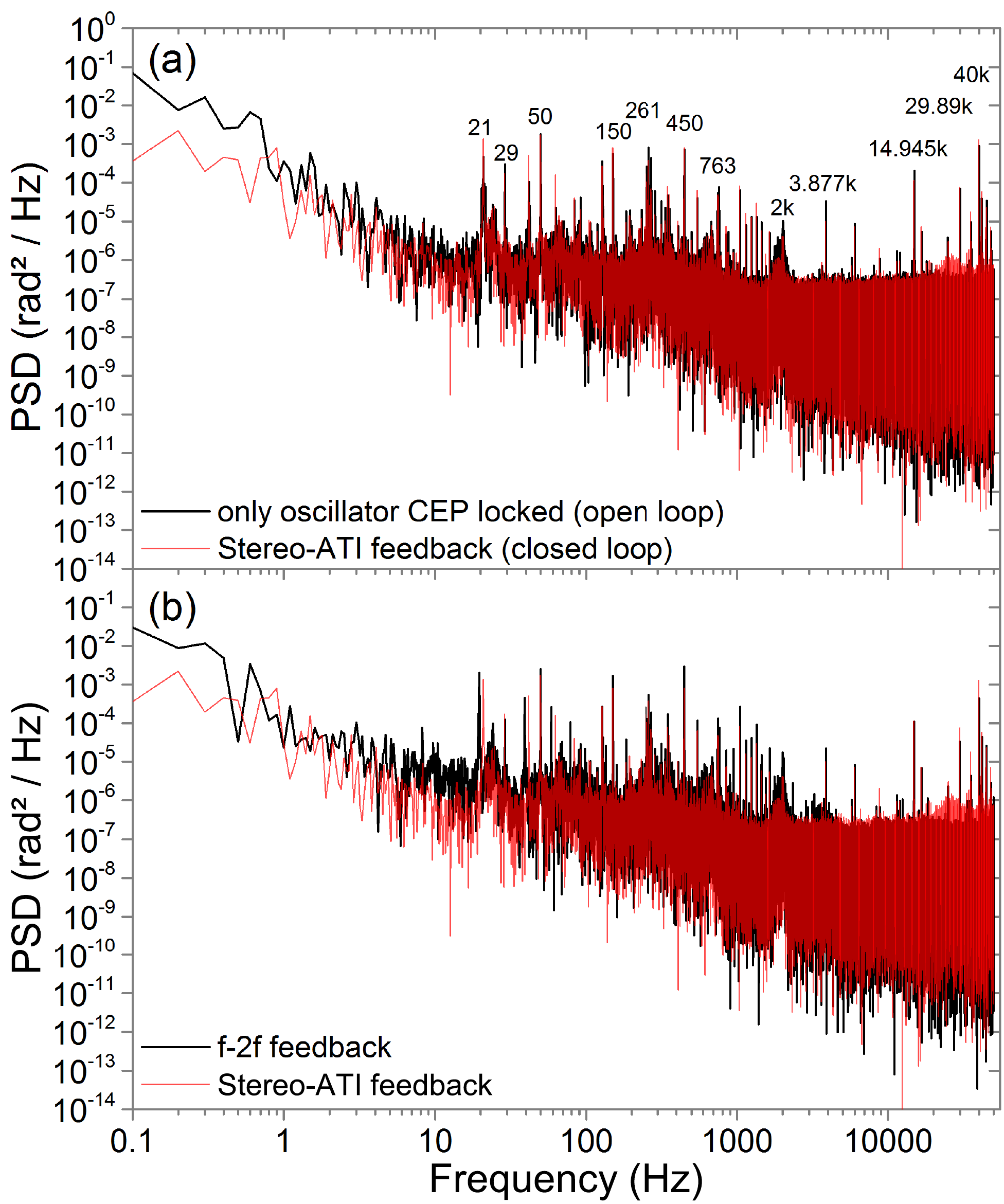}
\caption{Power spectral density from a Fourier analysis of 1M consecutive laser shots. 
Panel (a): comparison of open loop (oscillator CEP-locked only, black) and closed stereo-ATI 
feedback loop (red).
The main noise components are denoted.
The feedback with the stereo-ATI signal reduces the noise significantly over the full spectral 
range, except for a small region near 25\,kHz.
Panel (b): comparison of the f-2f feedback (black) and the stereo-ATI feedback (red).
}
\label{fig:FTA}
\end{figure}
Detecting the CEP of the 100\,kHz pulse train on a single shot basis, Fig. \ref{fig:CEPtrace}, 
allows for a frequency-resolved noise analysis of the laser system, which is shown in Figure 
\ref{fig:FTA}. 

It can be seen that for this OPCPA system, the majority of the noise remaining after locking the 
oscillator (black) lays below 1 kHz, and therefore a relatively slow feedback loop is sufficient to 
control most of the CEP noise of the amplified pulses. 
The feedback loop utilizing the stereo-ATI signal reduces the phase noise by several orders of 
magnitudes for low frequencies and lays well below the open loop result over the whole spectrum, 
as can be seen by comparing both traces in Fig. \ref{fig:FTA} (a). 
A few well-defined noise peaks and their harmonics are highlighted in Fig. \ref{fig:FTA} (a). 
Most of them correspond to acoustic noise in the laboratory.
Note, that between 20 and 30\,kHz the feedback from the stereo-ATI introduces additional noise, 
which is also seen in the CEP as a function of time at short time scales (Fig. \ref{fig:CEPtrace} 
(a, d, g)). 

Figure \ref{fig:FTA} (b) shows a comparisson of the frequency-resolved phase noise of the system in 
close-loop mode for the case of locking with the stereo-ATI signal (red) and the f-2f interferometer 
signal (black). 
The frequency analysis shows an improvement over almost the full spectral range when the stereo-ATI 
is utilized. 
Particularly in the range from about 1\,kHz to 4\,kHz the noise present with the f-2f-feedback is 
suppressed.
It is worth noticing that the data corresponding to the locking with the stereo-ATI signal is an 
in-loop measurement, while the f-2f locking case is measured out-of-loop. 
Nevertheless, these proof-of-principle results show the potential for the stereo-ATI to control the 
system at frequencies well over 10\,kHz. 
Future improvements to the PID feedback controls should allow for further reduction in noise and 
better measurement precision will lower the noise floor. 
In addition, the results show that the OPCPA system analyzed is well suited for performing 
CEP-sensitive experiments. 

In conclusion we have demonstrated the capability to measure and control the CEP of few-cycle laser 
pulses at repetition rates of 100\,kHz.  
This allowed us to characterize for the first time, the residual CEP noise from a high 
repetition rate OPCPA on a single-shot basis at the full repetition rate. 
This setup also enables the investigation of CEP-dependencies in ultrafast processes through the 
phase-tagging approach and facilitates highly efficient control of the CEP for steering processes 
like HHG driven by few-cycle laser pulses and in particular the generation of isolated attosecond 
XUV pulses. 
Due to the fast CEP evaluation ($\approx300$\,ns), the technique also allows for measurement and 
control of even higher repetition rate lasers up to 400\,kHz. 
Combined with the continuing progress in laser development \cite{Rothhardt2012, Prinz2015, 
Matyschok2013, Furch2017, Elu2017, Thire2017, Mero2015, Hadrich2016, Lavenu2017}, these features 
promise exciting developments in attosecond science.

\textbf{Funding.}
Horizon 2020 Framework Programme (H2020) (Laserlab-Europe EU-H2020 654148, ASPIRE 674960),
the DFG Grant PA 730/7 (SPP-1840 QUTIF).

\bibliography{default}
% \ifthenelse{\equal{\journalref}{ol}}{%
% \clearpage
% \bibliographyfullrefs{default}
% }{}
\end{document}